# First evidence of a strong Magneto-capacitance coupling at room temperature in integrated piezoelectric resonators

M. Maglione*, W. Zhu° & Z. H. Wang°

*Institute of Condensed Matter Chemistry of Bordeaux 1 ICMCB-CNRS 87, av. Dr.Schweitzer F-33608 PESSAC cedex France, ° Microelectronic Centre, School of Electrical and Electronic Engineering Nanyang Technological University, 50 Nanyang Avenue, Singapore 639798

Correspondence and requests for materials should be addressed to M.Maglione (e-mail: maglione@icmcb.u-bordeaux1.fr).

Abstract: In the vicinity of their resonance frequency, piezoelectric resonators are highly sensitive to small perturbations. The present report is focussed on the magnetic field as a perturbation source. First, magneto-dielectric modulation of more than 10% is achieved at room temperature on both ferroelectric single crystals and quartz discs. Since such piezoelectric resonators are now available as membranes directly integrated on Silicon wafer, we have checked the magneto-dielectric modulation in such resonators. We show here for the first time that a moderate magnetic field of $2.10^4$ Oersteds is able to efficiently tune the impedance of these resonators in their resonance window.

The modulation of dielectric materials through a magnetic excitation is currently attracting a lot of research work. Both the understanding of the magneto-dielectric effect and its potential applications are the driving forces for this interest. The direct interaction between an external magnetic field and the polarization of ferroelectric materials is one way to achieve such a magneto-dielectric effect. However, such an interaction is rather unusual and the number of magneto-ferroelectrics is limited[1]. Several breakthroughs towards magneto-dielectric modulation were recently achieved using multilayers[2,3] , ceramics[4] or single crystals[5,6]. However, the best magneto-dielectric modulation in these systems was only reported at cryogenic temperatures T<100K.

A very recent report evidenced the room temperature coexistence of ferroelectricity and ferromagnetic properties in nanostructured films made of $CoFe_2O_4$ imbedded in a $BaTiO_3$ matrix[7]. Furthermore, room temperature magneto-dielectric modulation of 1% under a static magnetic field of $10.10^4$ Oersteds was found in nanocomposites films made of $Bi_2O_3$-$Fe_2O_3$-$PbTiO_3$ [8]. This variation quickly decreased to 0.1% when the operating frequency used to measure the dielectric permittivity was swept from 1Hz to 100 kHz. It is thus highly desirable to look for improved systems displaying strong magneto-dielectric modulation.

A number of previous reports already pointed out that the interaction between an external magnetic field and a dielectric material may well be mediated by the piezoelectric distortion[2,3,6]. However, the processing of composites mixing magnetic and ferroelectric materials is not the only route to magneto-dielectric modulation. In the present paper, we show that the tuning of the operating frequency to the right range enables any purely piezoelectric material to be sensitive to an external magnetic field. Close to their piezoelectric resonance, the capacitance of such resonators can efficiently be tuned.

To probe this modulation, the resonators were hanged between two thin silver wires in the coil of an electro-magnet able to generate homogenous magnetic field between 0 and $2.10^4$ Oersteds. Small and soft wires are required in order for the resonator to be stress free and also to avoid any spurious interaction between the magnetic field and the wires. These wires were connected to the output port of a HP4194A impedance analyser. The main feature of this device for our purpose is that the operating frequency can be tuned with a great resolution between 100Hz and 15MHz. This allows us to scan any of the resonance frequency of standard resonators. The BaTiO$_3$ crystals were square plates of 4mmx4mm surface and 1mm thickness. The biggest surfaces were coated with sputtered gold layer acting as electrodes. We also probed the magneto-dielectric modulation of quartz resonators of disc shape. In both the BaTiO$_3$ plates and the quartz discs, the major faces were set perpendicular to the magnetic field which was homogenous throughout the whole surfaces.

The dielectric spectrum of a BaTiO$_3$ single crystal plate includes a large number of resonance modes with harmonics and satellites in the frequency window 50kHz-1.5MHz. This means that the BaTiO$_3$ resonators that are used here are by no way optimised as would be thin and long bars for example. At any of these resonances, the capacitance undergoes a strong S-shape curve (figure1a) and the dielectric losses show a sharp peak. From standard resonator models, these curves can be fitted by the real and imaginary part of a lorentzian respectively. What is of key relevance for the magneto-dielectric modulation is that the equivalent circuit of such resonators include an inductance part which is maximised when the operating frequency lies in the window where the capacitance has a negative slope[10]. The equivalent circuit emphasizing the inductive part L$_{em}$ stemming from the mechanical resonance is given in figure1b. This effective inductance is thought to be the source of the magnetic modulation of the capacitance and losses of the piezoelectric resonators in this frequency range. Indeed, the interaction between the external magnetic field and the inductance L$_{em}$ will lead to a shift of the resonance frequency and thus to a modulation of the effective capacitance close to the resonance. This appears clearly during the experiment which is displayed on figure 1a. In this, the 70kHz resonance of the BaTiO$_3$ plate was recorded at zero magnetic field and at $2.10^4$ Oersteds. The shift of the resonance under magnetic field is a strong evidence for the magneto-dielectric effect, the strength of which is 10% at the optimal frequency of 70.5kHz. Also the base capacitance is not affected by the magnetic field, confirming that direct magneto-dielectric modulation is hardly achieved in BaTiO$_3$ away from the resonance. This result is fully consistent with the already published report on quartz resonators[8] meaning that ferroelectricity is not the necessary condition for magneto-dielectric coupling to be efficient.

We next show that such magneto-dielectric coupling at the piezoelectric resonance can be extended to integrated piezoelectric resonators. As schematically shown on figure 2a, such membranes include a sol-gel processed PZT layer whose composition and microstructure has been optimised for the piezoelectric coupling to be the most efficient[11]. Under a dc bias of 20V and using a driving ac voltage of 30V peak-to-peak at a frequency of 143 kHz, a 2mm$^2$ membranes is able to be distorted by ~3µm. Using the same set up as the one used above for bulk piezoelectrics, we checked the influence of a dc magnetic field on the PZT membranes close to their resonance frequency. This is shown on figure 2b where the piezoelectric resonance of a 2x2 mm$^2$ resonator at 143kHz is clearly shifted to high frequencies under a magnetic field of $2.10^4$ Oersteds. For that particular experiment, the polarizing dc Bias was set to 30Volts while the probing ac signal was 1V. Such a 2% variation of C under the moderate magnetic field of $2.10^4$ Oersteds has never been reported up to now and we anticipate much stronger modulation in resonators of various shapes like cantilevers for example. Further experiments are underway to probe that. As to the model that could be drawn to fully understand, this magneto-capacitance modulation in integrated resonators, it will obviously be similar to the one used for bulk resonators. The very peculiar electrical equivalent circuit at the resonance frequency is thought to be a starting point for such a model.

To summarize, we have shown that the tuning of the operating frequency to the resonance range of a bulk piezoelectric resonator results in a 10% modulation of the dielectric parameters under a field of $2.10^4$ Oersteds. We have shown for the first time that this can be transferred to integrated PZT membranes which undergo a 2% modulation on the same conditions. This room temperature magneto-dielectric effect, which can be further optimised, opens the way to a number of applications in microelectronics, wireless communications and medical imaging. It is also proposed as an alternative route to the ongoing search for strong magneto-dielectric coupling at room temperature. Instead of looking for improved materials or architectures, the tuning of the operating frequency can be applied to very standard materials to tune their effective parameters.

**Figure Captions**

Figure 1: (a) lowest piezoelectric resonance mode of the BaTiO$_3$ resonator with and without a magnetic field of 2 10$^4$ Oersteds . As emphasized by the different markers, each of these curves has been reproduced several times showing the reversibility of the resonance shift. (b) Electrical equivalent circuit of a piezoelectric resonator emphasizing the electromechanical parameters (resistance R$_{em}$, inductance L$_{em}$ and capacitance C$_{em}$) and the out of resonance base parameters (capacitance C$_0$ and Resistance R$_0$).

Figure2: (a) schematic view of the piezoelectric membranes made of PZT sol-gel layers sandwiched between two Platinum electrodes and the necessary layers for the integration on the Silicon wafer. The most important technological step is the release of the Silicon wafer below the PZT resonator in order to free the membranes for z-type vibrations; (b) shift of the 143kHz resonance under a magnetic field of 2.10$^4$ Oersteds; the resonator was polarised through a dc bias of 40V and an ac signal of 1V pK-pK

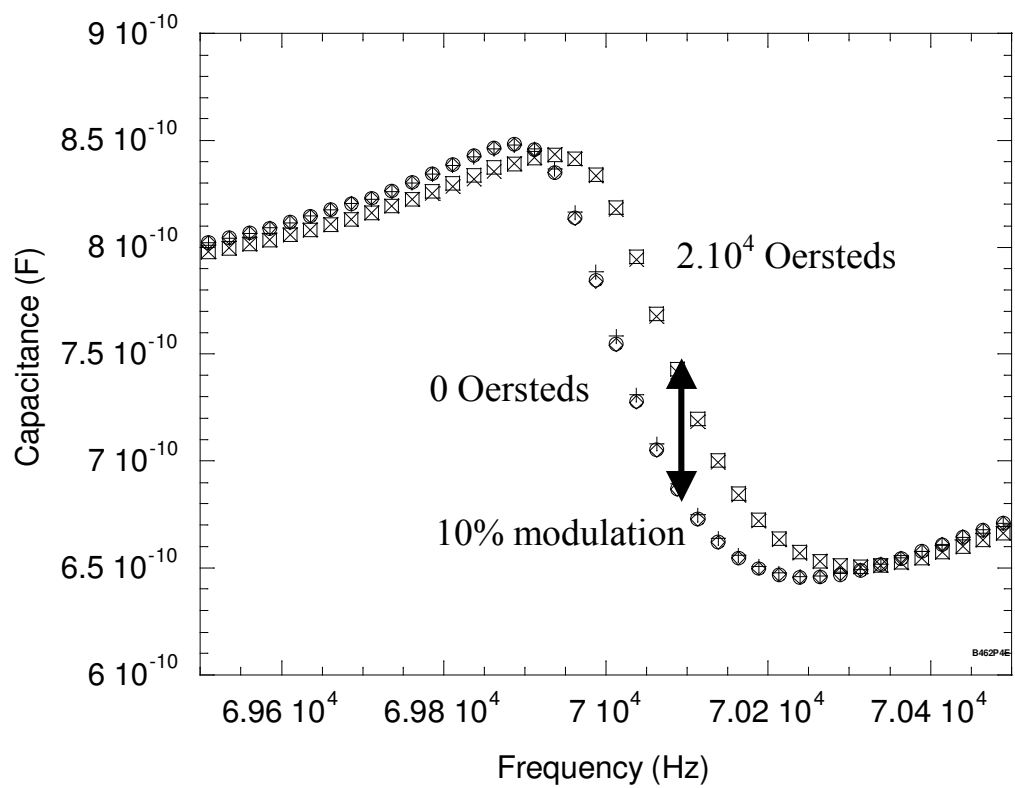

Figure 1a

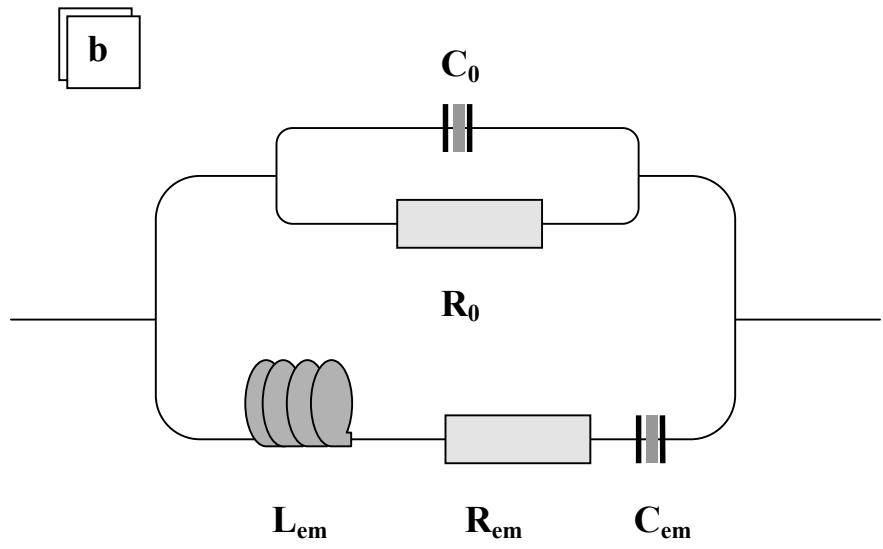

Figure 1b

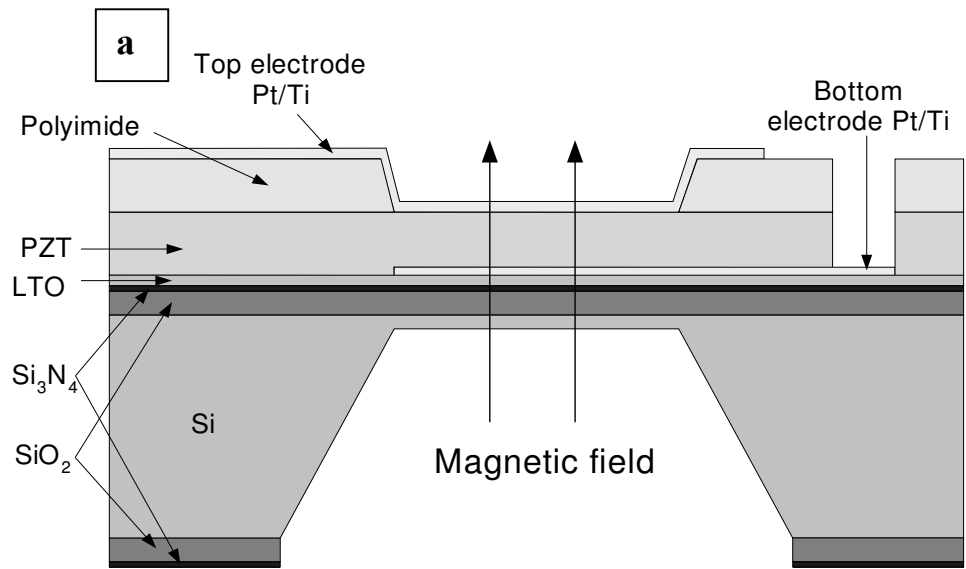

Figure 2a

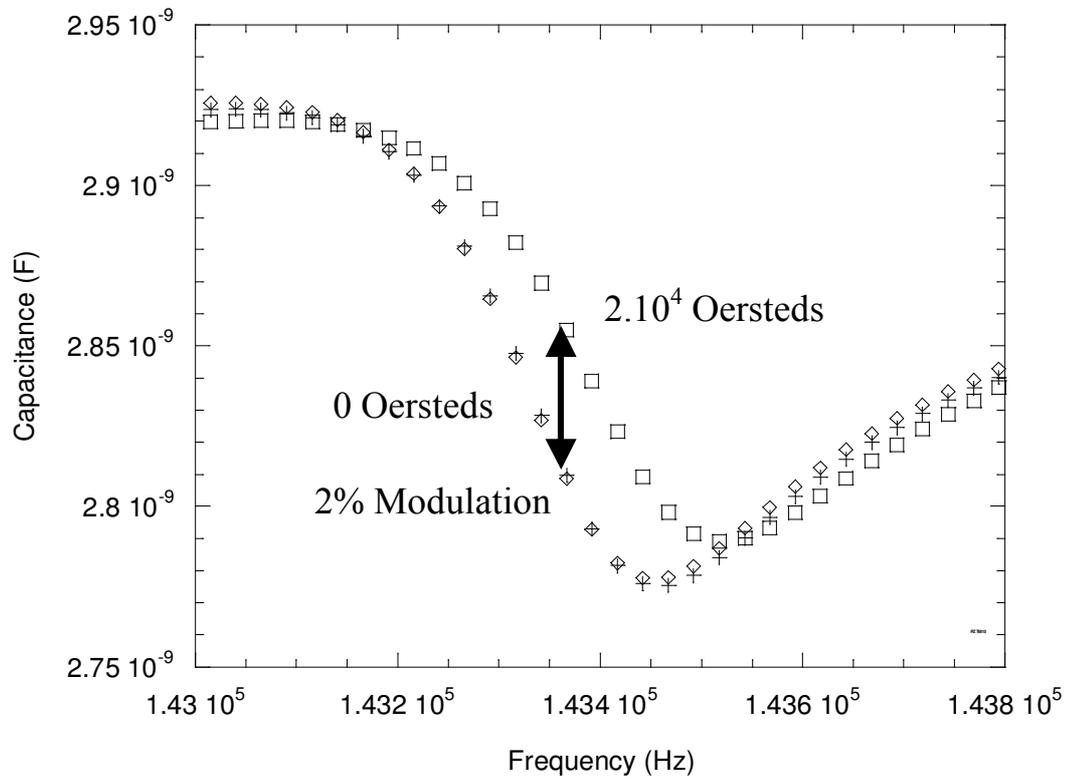

Figure 2b